\documentclass{iau} 
\usepackage{graphicx}
\usepackage{hyperref} 
\usepackage{xcolor}

\newcommand{\SB}{\texttt{SPHINCS\_BSSN }}
\newcommand{\Sb}{\texttt{SPHINCS\_BSSN}}
\newcommand{\Ma}{\texttt{MAGMA2 }}
\newcommand{\ma}{\texttt{MAGMA2}}
\def\be{\begin{equation}}
\def\ee{\end{equation}}
\def\bea{\begin{eqnarray}}
\def\eea{\end{eqnarray}}
\def\Msun{M$_{\odot}$ }

\title[Modelling astrophysical fluids with particles] 
{Modelling astrophysical fluids with particles}

\author[Stephan Rosswog]   
{Stephan Rosswog$^1$}

\affiliation{$^1$-
Astronomy and Oskar Klein Centre, Stockholm University,\\ AlbaNova, SE-10691 Stockholm, Sweden
\\ email: {\tt stephan.rosswog@astro.su.se} \\[\affilskip]}

\pubyear{2022}
\volume{362}  
\jname{Computational Astrophysics as a Discovery Tool}
\editors{A.C. Editor, B.D. Editor \& C.E. Editor, eds.}
\begin{document}

\maketitle

\begin{abstract}
Computational fluid dynamics is a crucial tool to theoretically explore the cosmos. In
the last decade, we have seen a substantial methodological diversification with a number of
cross-fertilizations between originally different methods. Here we focus on 
recent developments related to the Smoothed Particle Hydrodynamics (SPH) method. We briefly
summarize recent technical improvements in the SPH-approach itself, including
smoothing kernels, gradient calculations and dissipation steering.
These elements have been implemented in the Newtonian high-accuracy SPH code MAGMA2
and we demonstrate its performance in a number of challenging benchmark tests.
Taking it one step further, we have used these new ingredients also in the first particle-based,
general-relativistic fluid dynamics code  that solves the full set of Einstein equations, \Sb. We 
present the basic ideas and equations and demonstrate the code performance at examples 
of relativistic neutron stars that are evolved self-consistently together with the spacetime.

\end{abstract}

\firstsection
              
\section{Introduction}
A large fraction of the matter in the Universe can be modelled as fluids, which makes
computational gas dynamics a powerful tool in the theoretical exploration
of the Cosmos. While also  widespread in  engineering applications, {\em astrophysical}
gas dynamics comes with its own set of requirements and these sometimes trigger 
developments in new directions. Contrary to engineering applications, in astrophysics 
hard boundary conditions rarely play a role and often additional physical processes
beyond pure gas dynamics, e.g. magnetic fields, radiation or nuclear reactions, are main 
drivers of the evolution.
Gravity plays a central role in astrophysical gas dynamics. As a long range force, it can
easily accelerate gas to velocities that substantially exceed the local sound speed.
Therefore, shocks are ubiquitous in astrophysics, but they only occasionally play a role
in engineering applications. As a corollary, an astrophysical gas usually cannot --as in many engineering applications-- 
be treated as "incompressible", i.e. obeying a $\nabla \cdot \vec{v}=0$-condition, and
instead  the full set of compressible gas dynamics equations needs to be solved. A number of timely
astrophysical topics involve gas dynamics in curved spacetime, for example accretion flows 
around black holes or mergers of neutron stars.
\\
\begin{figure}{t}
 \centerline{\includegraphics[width=1\textwidth]{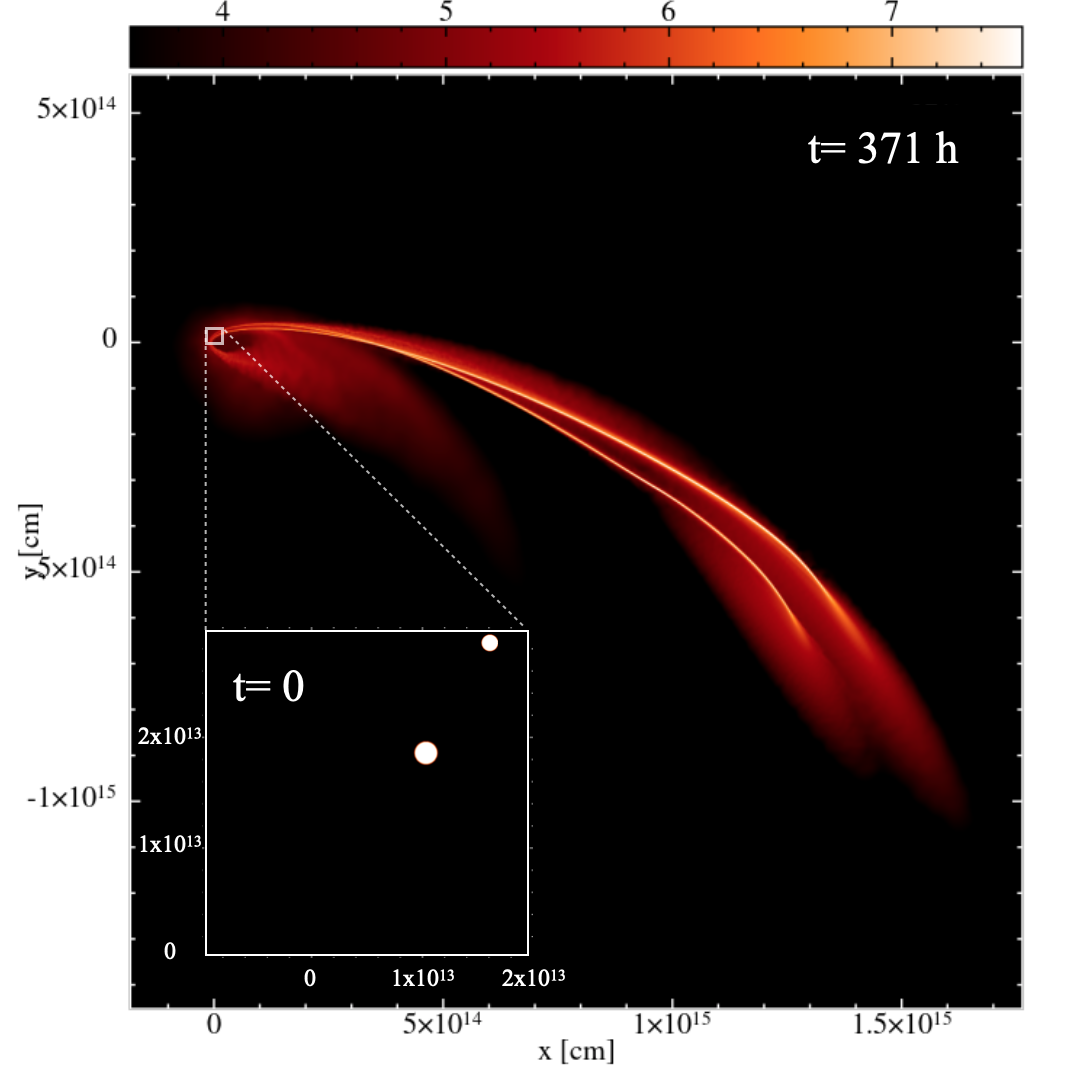}}
 \caption{Tidal disruption of a stellar binary system (67 and 36.8 M$_\odot$) by a $10^6$ M$_\odot$ 
 black hole (located at the origin; color-coded is the column density). About 10\% of all stellar tidal 
 disruption events could be disruptions of binary systems \cite[(Mandel \& Lewin 2015)]{mandel15}. The simulation has been
 performed with the code \Ma \cite[(Rosswog 2020)]{rosswog20a}.}
   \label{fig:DTDE}
\end{figure}
Even if gravity can be accurately treated in the physically rather simple Newtonian approximation,
its long-range nature makes it computationally very expensive and the resulting, often
filamentary gas structures can pose enormous methodological challenges with respect to
geometric adaptivity.  To illustrate this in an extreme example, we show in Fig.~\ref{fig:DTDE}
the tidal disruption of a stellar binary system (67 and 36.8 M$_\odot$) by a supermassive black hole 
($10^6$ M$_\odot$), located at the origin (simulation from \cite{rosswog20a}). Such binary
disruptions where actually both components become shredded have been found 
to make up a non-negligible fraction of all tidal disruption events (\cite[Mandel \& Levin 2015]{mandel15}).
The initial configuration of this simulation\footnote{The initial conditions for 
this simulation were kindly provided by I. Mandel and the corresponding stellar profiles by S. Justham.}
 is shown as inset in the lower left corner. Such a simulation with huge
changes in length and density scales, a complicated final geometry with the stars being stretched into extremely
thin gas streams that are held together by self-gravity and the majority of the "computational volume" being empty
(the initial stars cover $< 10^{-9}$ of the shown volume), are very serious computational challenges.
For such applications, particle methods have clear benefits: no additional computational infrastructure 
(such as an adaptive mesh) is needed, no computational resources are wasted on simulating the vast regions
of empty space and the particles simply move where the gas wants to flow. Moreover,  the excellent
advection properties of particle schemes allow to reliably follow the ejecta out to huge distances.\\
Methodologically, astrophysical gas dynamics was for a long time split into predominantly
Eulerian (mostly Finite Volume) and Lagrangian (mostly Smoothed Particle Hydrodynamics)
methods. But in the last decade computational methods have diversified, often combining
elements from different methods into "hybrids". One example of such a hybridization are so-called
"moving mesh methods" \cite[(Springel 2010; Duffell \& MacFadyen 2011; Duffell 2016; Ayache et al. 2022)]
{springel10,duffel11,duffell16,ayache22} where space is tessellated into  Voronoi-cells. Within 
these cells familiar Finite Volume techniques such as slope-limited reconstructions are applied
and at cell interfaces (either exact or approximate) Riemann solvers are used to determine
the numerical inter-cell fluxes. The cells themselves are often moved in a (quasi-)Lagrangian 
way, but, in principle, they can also be kept fixed in space or move with a velocity that is different 
from the local fluid velocity. In other words, these are "Adaptive-Lagrangian-Eulerian (ALE)" methods.
These methods inherit good shock capturing capabilities and are at the same time highly adaptive
and show good (though not perfect) numerical conservation properties.\\
Such ALE Finite Volume methods, however, are by no means restricted to {\em non-overlapping} 
Voronoi cells as basic geometric elements,  they can also be applied to freely moving, {\em overlapping} 
particles. This has been known in the numerical mathematics community for a long time (see e.g. \cite[Ben Moussa et al. 1999, 
Vila 1999, Hietel et al. 2000, Junk 2003]{benmoussa99,vila99,hietel00,junk03}), but has only found its way into astrophysics about a decade ago
(e.g. \cite[Gaburov \& Nitadory 2011; Hopkins 2015, Hubber et al. 2018]{gaburov11,hopkins15,hubber18}).\\
Our main focus here is on the Smoothed Particle Hydrodynamics (SPH) method and its recent developments. 
We aim at an improved SPH version that keeps  the robustness, geometric flexibility and excellent 
conservation properties of the original method, but is further improved in terms of accuracy. Even more ambitiously,
our goal is an accurate particle modelling of a relativistic fluid within a self-consistently evolving,
general relativistic spacetime. This goal has recently been reached \cite[(Rosswog \& Diener 2021)]{rosswog21a} after a 
string of new elements has been introduced which improve the accuracy of SPH \cite[(Rosswog 2010a, Cullen \& Dehnen 2010, 
Rosswog 2010b, Dehnen \& Aly 2012, Rosswog 2015b, Frontiere et al. 2017,  Rosswog 2020a,b)]{rosswog10a,cullen10,rosswog10b,dehnen12,rosswog15b,frontiere17,rosswog20a,rosswog20b}.
Most of these new elements are implemented into the Newtonian high-accuracy SPH code \Ma \cite[(Rosswog 2020a)]{rosswog20a}
which served also as a "test-engine" for many methodological experiments. The new elements have also found
their way into the first fully general relativistic, Lagrangian hydrodynamics code \SB \cite[(Rosswog \& Diener 2021)]{rosswog21a}. \\
This paper
is organized as flollows: in Sec.~\ref{sec:SPH} we discuss the recent improvements that have been
implemented into \Ma  and we demonstrate its performance in a number of challenging benchmark tests, in 
Sec.~\ref{sec:SPHINCS} we discuss the method and implementation of the first general relativistic SPH 
code that consistently solves the full set of Einstein equations and our results are finally  summarized 
in Sec.~\ref{sec:summary}.

\section{Recent improvements of Smoothed Particle Hydrodynamics}
\label{sec:SPH}
Here we briefly summarize frequently used SPH equations to set the stage for further improvements
and for a smooth transition to the relativistic case which will be described below.
A key ingredient of most SPH formulations is the density estimation at the position of a particle $a$
\be
\rho_a= \sum_b m_b W_{ab}(h_a).
\label{eq:rho_sum}
\ee
Here $m$ the particle mass, $W_{ab}(h_a)= W(|\vec{r}_a-\vec{r}_b|,h_a)$ is a smooth kernel function,
usually with compact support, and $h$ is the "smoothing length" 
which determines the support size of $W$. The usual approach is to keep each 
particle's mass constant in time so that exact mass conservation is ensured and no continuity 
equation needs to be solved. One can derive SPH equations elegantly 
from the SPH-discretized Lagrangian of an ideal fluid \cite[(e.g. Monaghan 2005)]{monaghan05}
\be
L= \sum_b m_b \left\{\frac{1}{2}v_b^2 -u(\rho_b,s_b) - \Phi_b\right\}.
\ee
Here, $v_b$ is the velocity of SPH particle $b$, $u_b$ its specific internal energy, $s_b$ its
specific entropy and $\Phi_b$ the gravitational potential. Applying the Euler-Lagrange equations  and 
the adiabatic form of the first law of thermodynamics, 
\be
\frac{d}{dt}\frac{\partial L}{\partial \vec{v}_a}- \frac{\partial L}{\partial \vec{r}_a}=0 \quad {\rm and} \quad \left(\frac{\p u}{\p \rho}\right)_a= \frac{P_a}{\rho_a^2},
\ee
yields the SPH momentum equation\footnote{Note that we are neglecting here small corrective terms, 
usually called "grad-h" terms, see Springel 2002, Monaghan 2002.}
\be
\frac{d\vec{v}_a}{dt}= - \sum_b m_b \left\{ \frac{P_a}{\rho_a^2} \nabla_a W_{ab}(h_a)+ \frac{P_b}{\rho_b^2} \nabla_a W_{ab}(h_b) \right\} + \vec{f}_{g,a}
\label{eq:N_Euler}
\ee
with $\vec{f}_g$ being the gravitational acceleration \cite[(Price \& Monaghan 2007)]{price07}.
A consistent energy evolution equation follows in a straight forward way by  translating
the first law of thermodynamics \cite[(see e.g. Rosswog 2009)]{rosswog09}
\be
\frac{du_a}{dt}= \frac{P_a}{\rho_a^2} \sum_b m_b \vec{v}_{ab} \cdot \nabla_a W_{ab}(h_a),
\label{eq:N_energy}
\ee
where $\vec{v}_{ab}= \vec{v}_a-\vec{v}_b$. 
It is worth mentioning that one has, of course, some freedom in the choice of variables and one can,
for example, also evolve the specific thermo-kinetic energy $\hat{e}= u + v^2/2$  according to
\be
\frac{d\hat{e}_a}{dt}= - \sum_b m_b \left\{{\frac{P_a \vec{v}_b}{\rho_a^2} \cdot \nabla_a W_{ab}(h_a) + \frac{P_b \vec{v}_a}{\rho_b^2} \cdot \nabla_a W_{ab}(h_b) } \right\}.
\label{eq:thermokinetic}
\ee
As will be seen later, this equation is very similar to the general relativistic evolution equation for the canonical energy
per baryon. For practical applications, the energy and momentum equations need to be enhanced by a 
mechanism to produce entropy in shocks, see Sec.~\ref{sec:AV}. 

\subsection{Kernel function}
\label{subsec:kernel}
The kernel function $W$ is a core ingredient of any SPH formulation. Traditionally cubic spline kernels
have been used \cite[(Monaghan 1992)]{monaghan92}, but they are of  moderate accuracy in density and gradient estimates (see e.g. Fig.~4 in \cite{rosswog15b})
and, for large neighbour numbers, they are prone to a "pairing instability", where particles begin to form pairs so that
resolution is effectively lost. A necessary condition for stability against pairing  is the non-negativity
the kernel's Fourier transform \cite[(Dehnen \& Aly 2012)]{dehnen12}. 
\cite{wendland95}  suggested a class of positively definite, radial basis functions of minimal
degree and these kernels are immune against the pairing instability.\\
After exhaustive experiments with various kernels \cite[(Rosswog 2015b)]{rosswog15b}, we settled for our \Ma code 
\cite[(Rosswog 2020a)]{rosswog20a} on a $C^6$-smooth Wendland kernel \cite[(Schaback \& Wendland 2006)]{schaback06} 
which overall delivered the best results. For experiments on {\em static}, perfect lattices other
high-order kernels actually delivered  density and gradient estimates of even higher accuracy, but in {\em dynamic} test cases the
Wendland kernel was by far superior. This is  because it maintains even in dynamical simulations a very 
regular particle distribution which is crucial for accurate kernel estimates. This kernel, however, needs a large number 
of neighbour particles in the SPH summations for accurate density and gradient estimates (see e.g. Figs. 4 and 5 in \cite{rosswog15b})
and so this improvement comes at some computational cost.
As a measure to keep the noise level very low, we choose in \Ma the smoothing length at each time step so that {\em exactly} 
300 neighbour particles contribute in the summations. Technically this is achieved via a very fast tree structure 
\cite[(Gafton \& Rosswog 2011)]{gafton11}, see \cite{rosswog20a}  for more technical details.

\subsection{Accurate gradients via matrix-inversion}
\label{subsec:gradients}
The standard approach in SPH is to represent the $\nabla P/\rho$-term on the RHS of the Euler equations via
expressions involving gradients of the kernel functions, as shown in Eqs.~(\ref{eq:N_Euler}) and (\ref{eq:N_energy}). This gradient
representation is anti-symmetric, $\nabla_a W_{ab}= - \nabla_b W_{ab}$, and therefore allows for a straight forward
enforcement of exact numerical conservation\footnote{Since one uses radial kernels, their gradients point in the directions
of the line joining two particles, this ensure angular momentum conservation. See e.g. Sec. 2.4 in Rosswog 2009b for a detailed discussion of conservation in SPH.}. 
While individual gradient estimates can be of moderate accuracy
only, the overall simulation may still have a high degree of accuracy since it strictly obeys Nature's conservation
laws.  It is important, though, that any potential improvement of gradient accuracy does not sacrifice one of SPH's 
most salient features, its excellent conservation properties.\\
Such an improvement is actually possible and one way to achieve it is by enforcing the exact reproduction of linear functions via a matrix 
inversion \cite[(Garcia-Senz et al. 2012)]{garciasenz12}. In the resulting gradient expression one term (that vanishes
for an ideal particle distribution) can be dropped and this omission guarantees the desired anti-symmetry 
of the gradient expression with respect to the exchanging $a \leftrightarrow b$. This gradient prescription
delivers gradient estimates that are several orders of magnitude more accurate than the standard SPH approach,
see Fig. 1 in \cite{rosswog15b}.
The new set of SPH equations uses the standard density calculation (\ref{eq:rho_sum}), but has  
momentum and energy equations modified according to\footnote{In the \Ma code paper we also explore an additional
SPH formulation that also uses this accurate gradient prescription.} 
\bea
\frac{d\vec{v}_a}{dt}&=& - \sum_b m_b \left\{ \frac{P_a}{\rho_a^2} \vec{G}_a +  \frac{P_b}{\rho_b^2} \vec{G}_b\right\}
\label{eq:momentum_IA},\\
\left(\frac{du_a}{dt}\right)&=& \frac{P_a}{\rho_a^2} \sum_b m_b \vec{v}_{ab} \cdot \vec{G}_a
 \label{eq:energy_IA},
\eea
where the gradient functions read
\be
\left(\vec{G}_{a}\right)^k= \sum_{d=1}^3 C^{kd}(\vec{r}_a,h_a) (\vec{r}_b - \vec{r}_a)^d W_{ab}(h_a)
\; \& \;
\left(\vec{G}_{b}\right)^k= \sum_{d=1}^3 C^{kd}(\vec{r}_b,h_b) (\vec{r}_b - \vec{r}_a)^d W_{ab}(h_b)
\label{eq:Gb}
\ee
and the "correction matrix" 
\be
\left(C^{ki} (\vec{r},h)\right)= \left( \sum_b \frac{m_b}{\rho_b} (\vec{r}_b - \vec{r})^k (\vec{r}_b - \vec{r})^i W(|\vec{r}-\vec{r}_b|,h)\right)^{-1}
\label{eq:corr_mat}
\ee
accounts for the local particle distribution. This equation set is based on {\em much} more accurate gradients,
but equally good at numerically conserving physically conserved quantities. This follows directly from the anti-symmetry
of the gradient functions with respect to the exchange $a \leftrightarrow b$ and is also confirmed practically in a simulation
of the violent collision between two main sequence stars (see Sec. 3.7.3  in \cite{rosswog20a}) which is usually
considered a worst-case scenario for energy conservation \cite[(Hernquist 1993)]{hernquist93}. In this test the conservation accuracy
of the matrix inversion approach is on par with the standard SPH formulation.

\subsection{Slope-limited reconstruction in the dissipative terms}
\label{sec:AV}
SPH has a reputation of being overly  dissipative, but the SPH equations as derived from the above 
Lagrangian do not involve any dissipation at all. Therefore, the applied dissipation is the responsibility of
the code developer. One way to add the dissipation that is needed in shocks is, as in typical Finite Volume 
methods, via Riemann solvers. This approach has occasionally been followed in SPH 
\cite[(Inutsuka 2002, Cha 2003, Cha 2010, Puri \& Ramachandran 2014)]{inutsuka02,cha03,cha10,puri14},
but more common is the use of artificial viscosity. While artificial viscosity is one of the oldest concepts
in computational fluid dynamics \cite[(von Neumann \& Richtmyer 1950)]{vonneumann50}, its modern 
forms are actually not that different from approximate Riemann solvers \cite[(e.g. Monaghan 1997)]{monaghan97}.\\
One way to add artificial viscosity (that is actually very close to the original suggestion of \cite{vonneumann50}
is to simply enhance the physical pressure $P$ by a viscous contribution $Q$, i.e. one replaces everywhere $P\rightarrow P+Q$
, where the viscous pressure is given by \cite[(Monaghan \& Gingold 1983; Frontiere et al. 2017)]{monaghan83,frontiere17}
\be
Q_a= \rho_a\left( -\alpha c_{s,a} \mu_a + \beta \mu_a^2 \right),
\label{eq:art_press}
\ee
and
\be
\mu_a= \rm{min} \left(0,\frac{\vec{v}_{ab} \cdot \vec{\eta}_a}{\eta_a^2 + \epsilon^2}\right),
\ee
$\vec{v}_{ab}= \vec{v}_a - \vec{v}_b$, $\vec{\eta}_a$ is the separation vector between particles $a$ and $b$, 
de-dimensionalized with particle $a$'s smoothing length, $\vec{\eta}_a= (\vec{r}_a-\vec{r}_b)/h_a$. The min-function 
ensures that the artificial pressure is only applied between approaching particles. The quantity $\mu_a$
is a measure of the "velocity jump" between the particles $a$ and $b$.
While this prescription works well in strong shocks, it can be more dissipative than actually needed,
especially when there is no shock at all.\\
This unwanted dissipation can be reduced in a similar way as in Finite Volume methods: rather than applying
the velocity jump calculated as the difference of the particle velocities $\vec{v}_a-\vec{v}_b$ (in Finite Volume language:
applying a zeroth-order reconstruction between the particles), one can 
perform a slope-limited reconstruction of the particle velocities from both the $a$- and the $b$-side to the
mid-point between the two particles and use  the jump between these reconstructed velocities at the midpoint
when calculating the quantity $\mu$. This different calculation of $\mu$ is the only change that is required, otherwise
the same equation (\ref{eq:art_press}) can be used. In \Ma we use a quadratic reconstruction together with a  van Leer slope limiter
\cite[(van Leer 1974; Frontiere et al. 2017)]{vanLeer74,frontiere17}, for the technical details we refer to our original code paper \cite[(Rosswog 2020a)]{rosswog20a}.
This reconstruction dramatically reduces unwanted dissipation even
if the dissipation parameters $\alpha$ and $\beta$ are kept at constant, large values. We found this effect 
particularly pronounced when simulating weakly triggered Kelvin-Helmholtz instabilities: without reconstruction
the instability growth was effectively suppressed (as seen in more traditional SPH approaches) while {\em with} 
reconstruction the instability grows even at low resolution and with large, constant dissipation parameters $\alpha$ and $\beta$
at a rate very close to the expected one  (see Fig.~20 in \cite{rosswog20a}).

\subsection{Steering dissipation by entropy monitoring}
One can go even one step further in reducing dissipation: in addition to the above described slope-limited reconstruction,
one can also make the dissipation parameters $\alpha$ and $\beta$ in Eq.(\ref{eq:art_press})  time-dependent 
\cite[(Morris \& Monaghan 1997; Rosswog et al. 2000; Cullen \& Dehnen 20210; Rosswog 2015b)]{morris97,rosswog00,cullen10,rosswog15b}. Following \cite{cullen10}, we calculate  in \Ma at each time step for each particle
a desired value $\alpha^{\rm des}$ for the dissipation parameter  $\alpha$  (we use $\beta=2 \alpha$). If the  current value 
at a particle $a$, $\alpha_a$, is larger than  $\alpha^{\rm des}_a$, we let it decay exponentially according to 
\be
\frac{d\alpha_a}{dt}= - \frac{\alpha_a - \alpha_0}{30 \tau_a},
\ee
where $\tau_a= h_a/c_{s,a}$ is the particle's dynamical time scale and $\alpha_0$ is a low floor value (which can be zero). Otherwise, if $\alpha^{\rm des}_a > \alpha_a$, 
the value of $\alpha_a$  is instantaneously raised to $\alpha^{\rm des}_a$.\\
The novel part in our prescription \cite[(Rosswog 2020b)]{rosswog20b}  is how we determine $\alpha^{\rm des}$ or, in other words, how we 
determine the exact amount of needed dissipation. The main idea is that we are simulating an ideal fluid  which should 
conserve entropy exactly. In our approach exact entropy conservation is not enforced, so we can monitor it at each particle and use its
degree of non-conservation to steer dissipation. A non-conservation of entropy can be the result of a passing shock, or
to a much lower extent, it can result from particles becoming noisy for purely numerical reasons. In both cases
some amount of dissipation should be applied. For now, we use $s= P/\rho^\Gamma$ as an entropy measure and 
we monitor over each time step $\Delta t$ the relative entropy violation $\dot{\xi}_a\equiv(\Delta s_a/\Delta t)/(s_a/\tau_a)$. By 
numerical experiments we determine
a relative entropy violation that can be tolerated without need for dissipation, $\dot{\xi}_0$, and a value $\dot{\xi}_1$, where full
dissipation should be applied, in between we smoothly increase the dissipation values, see \cite{rosswog20b}  for the technical details. 
This way of steering the dissipation parameter has been found to work very well, it robustly switches on 
in shocks, but only leads to very low dissipation values otherwise. As an example, we show a Rayleigh-Taylor instability in
Fig.~\ref{fig:RaTa}, where the density (left panel) evolves in very close agreement with literature results (e.g. Frontiere et al. 2017).
Non-negligible amounts of dissipation are only triggered in the direct interface between the initial high- and low-density fluid (right panel),
elsewhere the dissipation is essentially zero.

\subsection{MAGMA2 results}
\begin{figure}[h]
\centerline{\includegraphics[width=1.1\textwidth,angle=0]{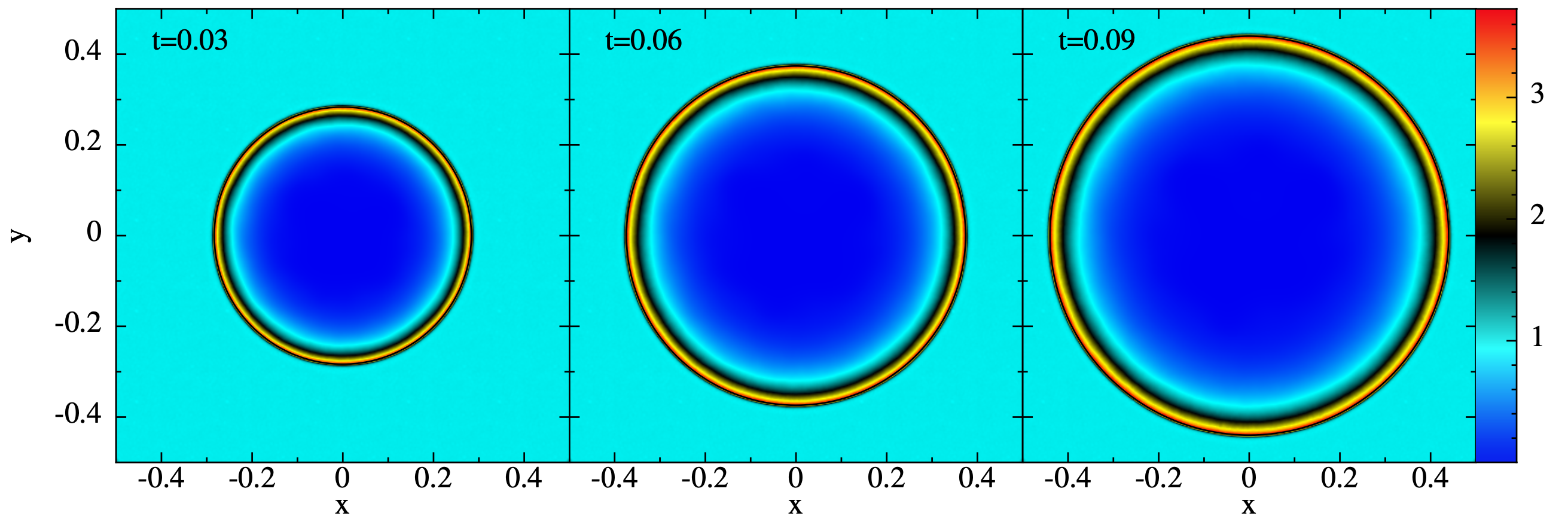}}
\caption{Density evolution in a Sedov-Taylor blast wave. The outermost black ring at the leading edge of the blast is the over-plotted exact solution. Simulation performed with code \ma.}
\label{fig:Sedov}
\end{figure}
Here we only show  a few tests: a Sedov-Taylor explosion as an example for a shock, a Rayleigh-Taylor instability 
as an instability example and two Schulz-Rinne tests as examples of complex shock-vortex interactions. 
For a Kelvin-Helmholtz test (density and triggered dissipation) we refer to
\href{http://compact-merger.astro.su.se/Movies1/KH1024_rho_AV_v5.mov}{\color{blue}a movie on the author's website}. For more 
tests and the technical details of a number of benchmark tests we refer to Rosswog (2020a,b).\\

{\em Sedov blast wave}\\
A classic, multi-dimensional shock problem is the Sedov-Taylor explosion test where
a strong, initially point-like blast expands into a low density environment \cite[(Sedov 1959; Taylor 1950)]{sedov59,taylor50}. 
For a given explosion energy $E$, an ambient medium density $\rho$ and polytropic $\Gamma=5/3$, 
the blast wave radius propagates according to $r(t)\approx 1.15 [(E t^2)/\rho]^{1/5}$ and the density
jumps by the strong-explosion limit factor $(\Gamma+1)/(\Gamma-1)= 4$. Behind the shock the density drops 
quickly and finally vanishes at the centre of the explosion. We show in Fig.~\ref{fig:Sedov} a cut through the
3D density as a function  of time. Also shown (as leading black circle), but hardly visible, is the exact solution,
which demonstrates the accurate agreement between our numerical and the exact solution. No deviation from
perfect spherical symmetry is visible and also the particle values are (practically noise-free) lying on top
of the exact solution, see Fig.~10 in Rosswog (2020a).\\

\begin{figure}[h]
\centerline{\includegraphics[width=0.8\textwidth,angle=0]{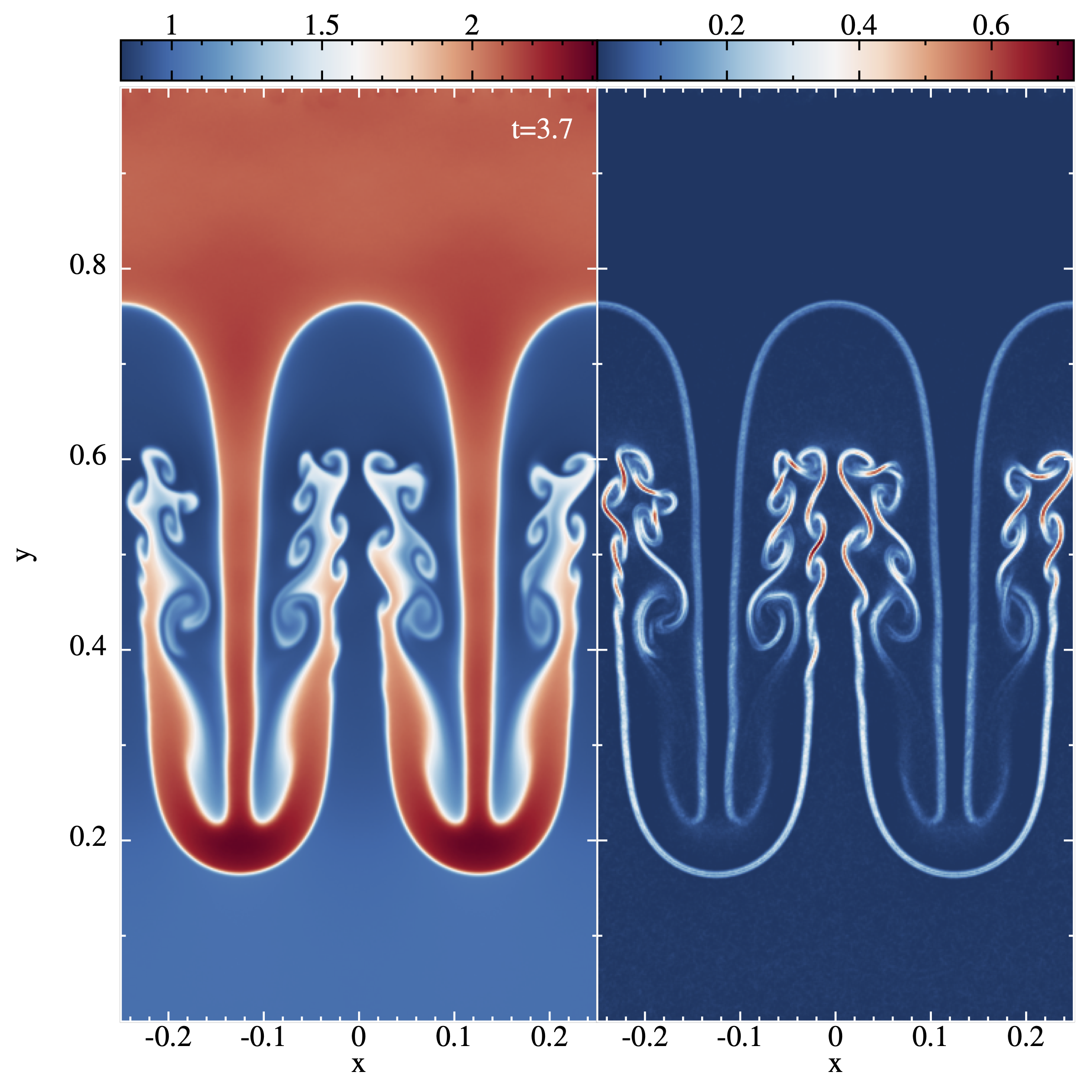}}
\caption{Shown are the density (left) and the applied dissipation parameter $\alpha$ ($\beta=2 \alpha$)
as steered via monitoring the local entropy conservation (right) in a 
Rayleigh-Taylor instability. Note that dissipation is nearly entirely absent throughout most of the flow.
Simulation performed with code \ma.}
\label{fig:RaTa}
\end{figure}
{\em Rayleigh-Taylor test}\\
The Rayleigh-Taylor instability is a standard probe of the subsonic growth of a small perturbation. 
In its simplest form, a layer of higher density rests on top of a layer with lower density  in a 
constant acceleration field, e.g. due to gravity. While the denser fluid sinks down, it develops a 
characteristic, "mushroom-like" pattern. Simulations with traditional SPH implementations have 
shown only retarded growth or even a complete suppression of the instability \cite[(Abel 2011; 
Saitoh \& Makino 2013)]{abel11,saitoh13}. We set up this test case as Frontiere et al. (2017) and our \Ma results
show a healthy growth of the instability, see Fig.~\ref{fig:RaTa}, left panel. Note in particular, that our entropy 
steering triggers dissipation in only a very limited region of space, while the bulk of the simulated
volume has essentially no dissipation (right panel).\\

\begin{figure}
\begin{center}
\centerline{\includegraphics[width=1.1\textwidth,angle=0]{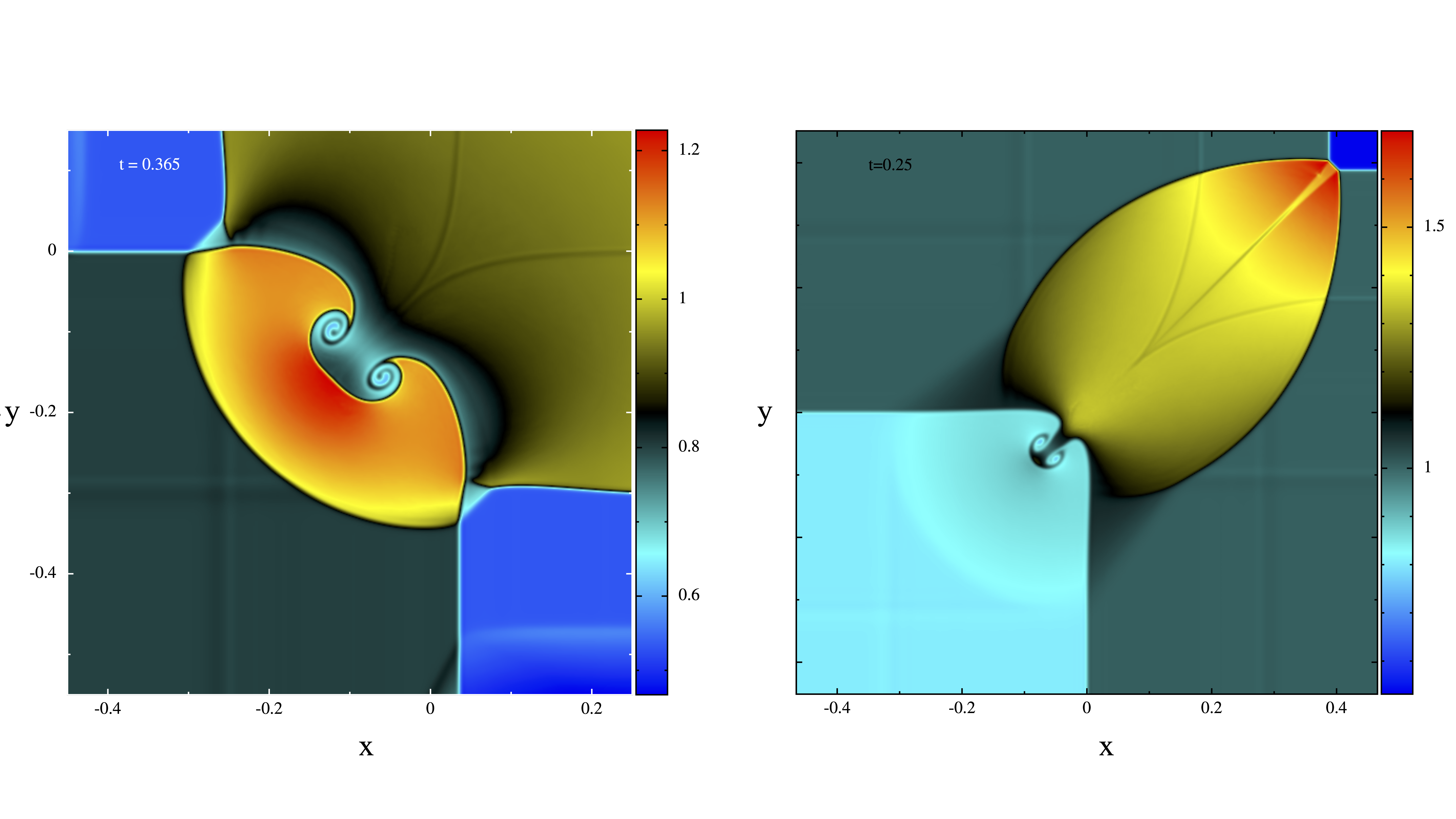}}
\vspace*{-0.7cm}
 \caption{Schulz-Rinne tests. In these challenging tests initially four constant states meet in one corner
 of the $xy$-plane. In the subsequent evolution complex shocks and vortex structures form. Our \Ma results
 are in close agreement with those found in Eulerian simulations.}
   \label{fig:SchulzRinne}
\end{center}
\end{figure}

{\em Schultz-Rinne tests}\\
\cite{schulzrinne93}  designed a particularly challenging set of tests in which initially four constant states
meet in one corner and the initial values are chosen so that one elementary wave, either a shock, a 
rarefaction or a contact discontinuity appears at each interface. During the evolution, complex flow patterns
emerge, involving shocks and vorticity, for which no exact solutions are known. These tests are considered challenging 
benchmarks for multi-dimensional hydrodynamics codes \cite[(Schulz-Rinne 1993; Lax \& Liu 1998; 
Kurganov \& Tadmor 2002; Liska \& Wendroff 2003)]{schulzrinne93,lax98,kurganov02,liska03} and we are only 
aware of one study that tries to
tackle these tests with (a Riemann solver version of) SPH  (Puri  \& Ramachandran 2014), with mixed success. 
We show in Fig.~\ref{fig:SchulzRinne}
two such tests (produced with the 3D code; 10 particle layers in $z$- and 660$\times$660 particles
in $xy$-direction; polytropic $\Gamma=1.4$), for further examples we refer to the \Ma code paper 
(Rosswog 2020a).  Fig.~\ref{fig:SchulzRinne} shows crisp and  noise-free mushroom-like
structures that are in very good agreement with the Eulerian results that can be found in the literature 
(e.g. Lax \& Liu 1998; Liska \& Wendroff 2003). 

\section{Smoothed Particle Hydrodynamics In Curved Spacetime: the SPHINCS\_BSSN code}
\label{sec:SPHINCS}
Motivated by the splendid prospects of multi-messenger astrophysics \cite[(Rosswog 2015a, Abbot et al. 2017, 
Barack et al. 2019, Kalogera et al. 2021)]{rosswog15a,abbott17,barak19,kalogera21}
our ultimate goal is to develop a general relativistic hydrodynamics code that consistently solves
for the evolution of  spacetime, but models the fluid with particles. We expect that a particle
method has clear benefits (compared to the current Eulerian approaches) in following the small amounts 
of ejecta, $\sim 1$\% of the binary mass, that are responsible for the entire electromagnetic display of 
a compact binary merger. \\
Since the general relativistic evolution of spacetime is a hyperbolic problem, 
we need to integrate to spacetime geometry forward in time, while in the Newtonian approach 
(with an infinite propagation speed of gravity) we solve an elliptic problem where the gravitational forces 
are calculated from  the instantaneous matter state.
The methods to evolve spacetime have substantially matured in the last two decades and can be found
in recent textbooks on Numerical Relativity (Alcubierre 2008, 
Baumgarte \& Shapiro 2010, Rezzolla \& Zanotti 2013, Shibata 2016, Baumgarte \& Shapiro 2021), we 
 decided to follow
the well-established BSSN-approach for evolving the spacetime on a computational mesh,
very similar to what is done in Eulerian approaches, but to evolve matter via Lagrangian particles.
This strategy is implemented in our newly developed Numerical Relativity code, \SB (Rosswog \& Diener 2021).

\subsection{General-relativistic hydrodynamics}
General relativistic SPH equations can be derived  similarly to the Newtonian approach \cite[(e.g. Monaghan \& Price 2001),]{monaghan01} 
see Rosswog (2009), Sec. 4.2 for a step-by-step derivation of the equations that we will use.
Instead of discretizing the gas into particles of constant mass, one now assigns to each particle a baryon number
$\nu_b$ that remains a constant-in-time property of each particle. One chooses a "computing frame" in which the
simulations are performed and calculates a computing frame baryon number density at the position of a particle
 $a$ according to
\be
N_a= \sum_b \nu_b\, W(|\vec{r_a} - \vec{r}_b|,h_a),
\label{eq:N_sum}
\ee
where $W$ is an SPH smoothing kernel (we use the same $C^6$-smooth Wendland kernel as in \ma). 
In other words, we calculate the density $N$ just as the mass
density in Newtonian SPH, Eq.~(\ref{eq:rho_sum}), but with particle masses being replaced with baryon numbers. The
number density in the local rest frame density of a particle, $n$, is related to $N$ via
\be
N= \sqrt{-g}\, \Theta n,
\label{eq:N_def}
\ee
where $g$ is the determinant of the spacetime metric and the generalized Lorentz factor is
given by
\be
\Theta\equiv \frac{1}{\sqrt{-g_{\mu\nu} v^\mu v^\nu}} \quad {\rm with} \quad v^\alpha=\frac{dx^\alpha}{dt}.
\label{eq:theta_def}
\ee
Similar to the Newtonian approach, one can start from a discretized Lagrangian of an ideal fluid
\be
L= -\sum_b \nu_b \left(\frac{1+u}{\Theta}\right)_b.
\ee
Note that we have followed here the convention that we measure all energies
in units of the baryon rest mass $m_0 c^2$.
We base our numerical evolution variables on the canonical momentum per baryon and
the canonical energy per baryon as they follow from the above Lagrangian.
The canonical momentum per baryon reads
\be
(S_i)_a = (\Theta \mathcal{E} v_i)_a,
\label{eq:can_mom}
\ee
where $\mathcal{E}= 1 + u + P/n$ is the relativistic enthalpy per baryon and the
canonical energy per baryon
\be
e_a= \left(S_i v^i + \frac{1 + u}{\Theta}\right)_a = \left(\Theta \mathcal{E} v_i v^i + \frac{1 + u}{\Theta}\right)_a.
\label{eq:can_en}
\ee
These quantities are evolved in time according to 
\be
\frac{d(S_i)_a}{dt}= -\sum_b \nu_b \left\{ \frac{P_a \sqrt{-g_a} }{N_a^2}  \;  \frac{\p W_{ab}(h_a)}{\p x_a^i}
+  \frac{P_b \sqrt{-g_b}}{N_b^2} 
 \; \frac{\p W_{ab}(h_b)}{\p x_a^i}
\right\}
+ \left(\frac{\sqrt{-g}}{2N} T^{\mu \nu} \frac{\p g_{\mu \nu}}{\p x^i}\right)_a
\label{eq:dSdt}
\ee
and 
\be
\frac{d e_a}{dt} = -\sum_b \nu_b \left\{ \frac{P_a \sqrt{-g_a}}{N_a^2}  \;  v_b^i   \;  \frac{\p W_{ab}(h_a)}{\p x_a^i} +  
\frac{P_b \sqrt{-g_b}}{N_b^2} \;  v_a^i \; \frac{\p W_{ab}(h_b)}{\p x_a^i} \right\} -\left(\frac{\sqrt{-g}}{2N} T^{\mu \nu} \frac{\p g_{\mu \nu}}{\p t}\right)_a.
\label{eq:dedt}
\ee
Note that our equations for the conservation of baryon number, Eq.~(\ref{eq:N_sum}), momentum, Eq.~(\ref{eq:dSdt}), and 
Eq.~(\ref{eq:dedt}) (compare to Eq.(\ref{eq:thermokinetic})), have a very "Newtonian look and feel". But while they are very 
convenient for the numerical evolution, they are actually not the physical variables that we are really interested in, these 
are $n,v^i$ and $u$. This means that we have to recover the physical variables $n,v^i, u$ at every time step from the 
numerical variables $N, S_i$ and $e$. But this is a price that also  Eulerian approaches have to pay, and we recover 
the physical variables with very similar methods, see Sec.2.2.4 in Rosswog \& Diener (2021) for the technical details.
We also need to add dissipative terms in \SB and we follow a  strategy similar to the one used in \ma:
a) we apply a slope limited reconstruction in the dissipative terms and b) we steer the dissipation by monitoring the entropy
change at every particle and time step. The details can be found in Sec. 2.2.3 of Rosswog \& Diener (2021).
 
\subsection{Evolving the spacetime via the BSSN formulation} 
To robustly evolve the spacetime, we have implemented two frequently used 
variants of the BSSN equations in \Sb, the ``$\Phi$-method''
\cite[(Nakamura et al. (1987), Shibata \& Nakamura (1995), Baumgarte \& Shapiro (1999))]{nakamura87,shibata95,baumgarte99} and the ``$W$-method''
\cite[(Tichy \& Marronetti (2007), Marronetti et al. (2008))]{tichy07,marronetti08}. The
complete set of BSSN equations is very lengthy and will therefore not be reproduced here.
It is described in detail in a number of Numerical Relativity textbooks \cite[(Alcubierre 2008, 
Baumgarte \& Shapiro 2010, Rezzolla \& Zanotti 2013, Shibata 2016, Baumgarte \& Shapiro 2021)]{alcubierre08,baumgarte10,rezzolla13,shibata16,baumgarte21} 
and can also be found in~\cite{rosswog21a}. For all the tests presented here we use the ``$\Phi$-method''.

\subsection{Coupling between fluid and spacetime}
As can be seen from the hydrodynamic equations Eqs.(\ref{eq:dSdt}) and (\ref{eq:dedt}), the fluid 
needs the derivatives of the metric (known on the mesh) at each particle position. The evolution of
the metric (evolved on a mesh), in turn, is governed by the energy momentum tensor $T_{\mu\nu}$ that is known at the particle
positions. We therefore need to continuously map $T_{\mu\nu}$ from the particles to the mesh ("P2M-step")
and $\partial_\lambda g_{\mu\nu}$ from the mesh to the particles ("M2P-step").\\
 In the P2M-step we have 
experimented with methods that are frequently used in particle-mesh methods \cite[(Hockney \& Eastwood 1988)]{hockney88}
and with common SPH kernels. But a general relativistic self-gravitating system is numerically very delicate
and we did not find these methods accurate enough for our purposes. For example, an initial neutron star setup according to
a Tolman-Oppenheimer-Volkoff solution, did not stay close to its equilibrium solution. We found {\em much} better solutions
when using kernels that are frequently used in the context of "vortex methods" (Cottet \& Koumoutsakos 2000).
These kernels are very accurate for close to uniform particle distributions, but they are not  positive definite
and they require the cancellation of positive and negative contributions. If applied naively everywhere, this can lead to Gibbs-phenomena-like
spurious oscillations near the surface of the star. Therefore we have implemented a hierarchy of kernels
of decreasing order with only the least accurate, "parachute" kernel being strictly positive definite. Applying this hierarchy of kernels
led to very good results, for details of this approach, we refer to Sec. 2.4 in Rosswog \& Diener (2021).\\
The M2P-step turned out to be less delicate, here we use a quintic Hermite polynomial in generalization
of the procedure described in \cite{timmes00}  to ensure that the 
interpolated values are $C^2$ when a particle passes from one grid cell to another.

\begin{figure}
\begin{center}
\centerline{\includegraphics[width=\textwidth,angle=0]{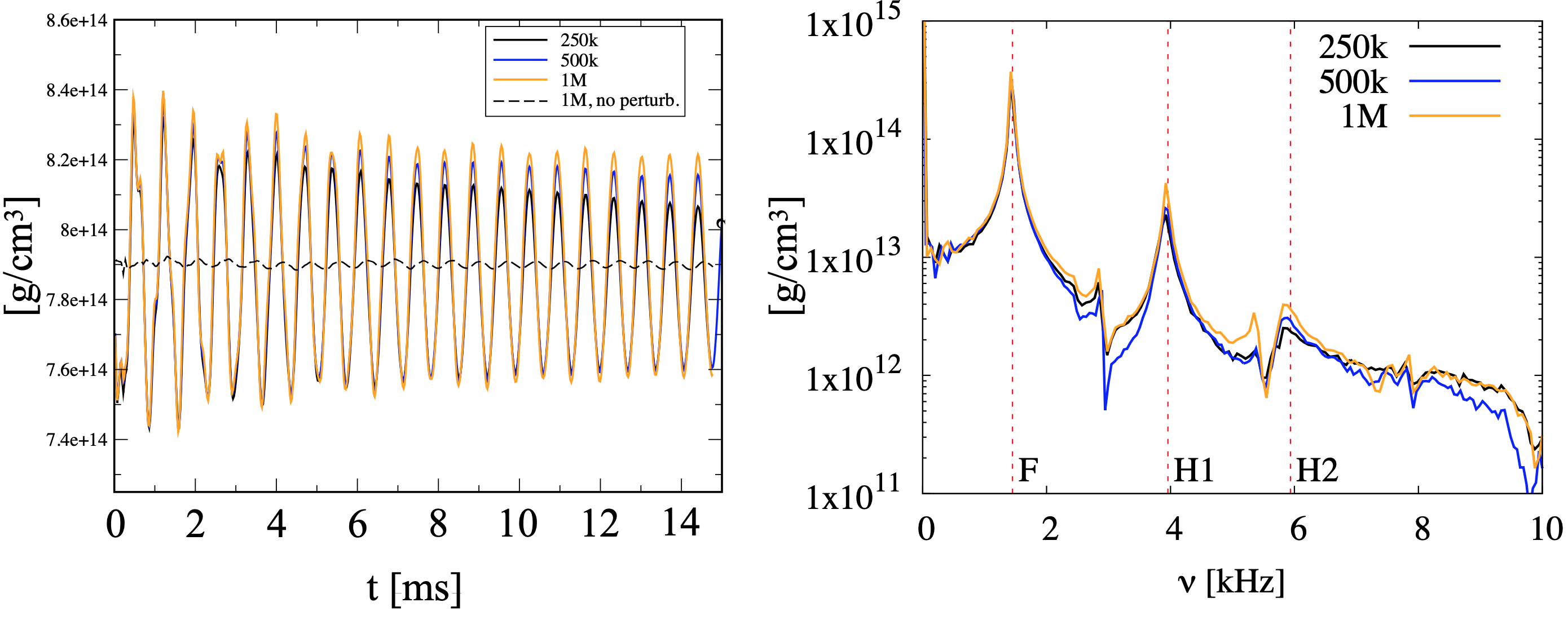}}
\vspace*{-0cm}
 \caption{Oscillations of a perturbed neutron star ($\Gamma=2.0$). Left: central density as function of time. Right: Fourier spectrum of the central density oscillations,
  the red dashed vertical lines are the fundamental normal mode frequency (F) and the next two higher mode frequencies (H1, H2) as determined by Font et al.(2002). Simulations performed with code \Sb.}
   \label{fig:NS_osc}
\end{center}
\end{figure}

\subsection{\texttt{SPHINCS\_BSSN} results}
Our full evolution code has been scrutinized in a number of standard test cases such as relativistic
shock tubes (to test special relativistic hydrodynamics), oscillations of neutron stars in a frozen
spacetime ("Cowling approximation"; to test general relativistic hydrodynamics), oscillations of neutron 
stars when the spacetime is dynamically evolved (to test the combined hydrodynamic-plus-spacetime evolution) 
and last, but not least, the challenging "migration test". In this test, a neutron star is prepared on the unstable 
branch and migrates, depending on the type of perturbation, either via violent oscillations onto the stable branch, 
or collapses into a black hole. In the following, we will only describe the fully relativistic, oscillating neutron star 
and the migration test. We use units in which  $G=c=1$ and masses are measured in solar units. For the 
other tests and more details we refer to the original paper (Rosswog \& Diener 2021), a first set of neutron star 
merger simulations with \texttt{SPHINCS\_BSSN} can be found in \cite{diener22a}.\\

{\em Oscillating neutron star in a dynamical spacetime}\\
In this test, we set up a 1.40 \Msun (gravitational) neutron star, modelled with a polytropic equation of state ($P= K n^\Gamma; K=100 {\rm \; and \;} \Gamma= 2.0$;
keep in mind our convention of measuring energies in $m_0c^2$),
according to the corresponding Tolman-Oppenheimer-Volkoff (TOV) solution. Subsequently, the star receives a small, radial velocity perturbation and
is evolved in its dynamical spacetime. The resulting central density evolution is shown in the left panel of Fig.~\ref{fig:NS_osc} (for 250 000, 500 000 and 
1 million SPH particles). The unperturbed stars stay close to but slightly oscillate (due to truncation error) around the  TOV solution (black line). 
We perturb the stars and measure their oscillation frequencies. The  fundamental normal mode (F: 2.696 kHz) and 
the first two overtones (H1: 4.534 kHz, H2: 6.346 kHz) as determined by Font et al. (2002) via a 3D Eulerian high resolution shock capturing code
are shown as the red dashed lines in the right panel. We find excellent agreement of the oscillation frequencies at the $\sim1\%$ level.\\
\begin{figure}
\begin{center}
\centerline{\includegraphics[width=\textwidth,angle=-90]{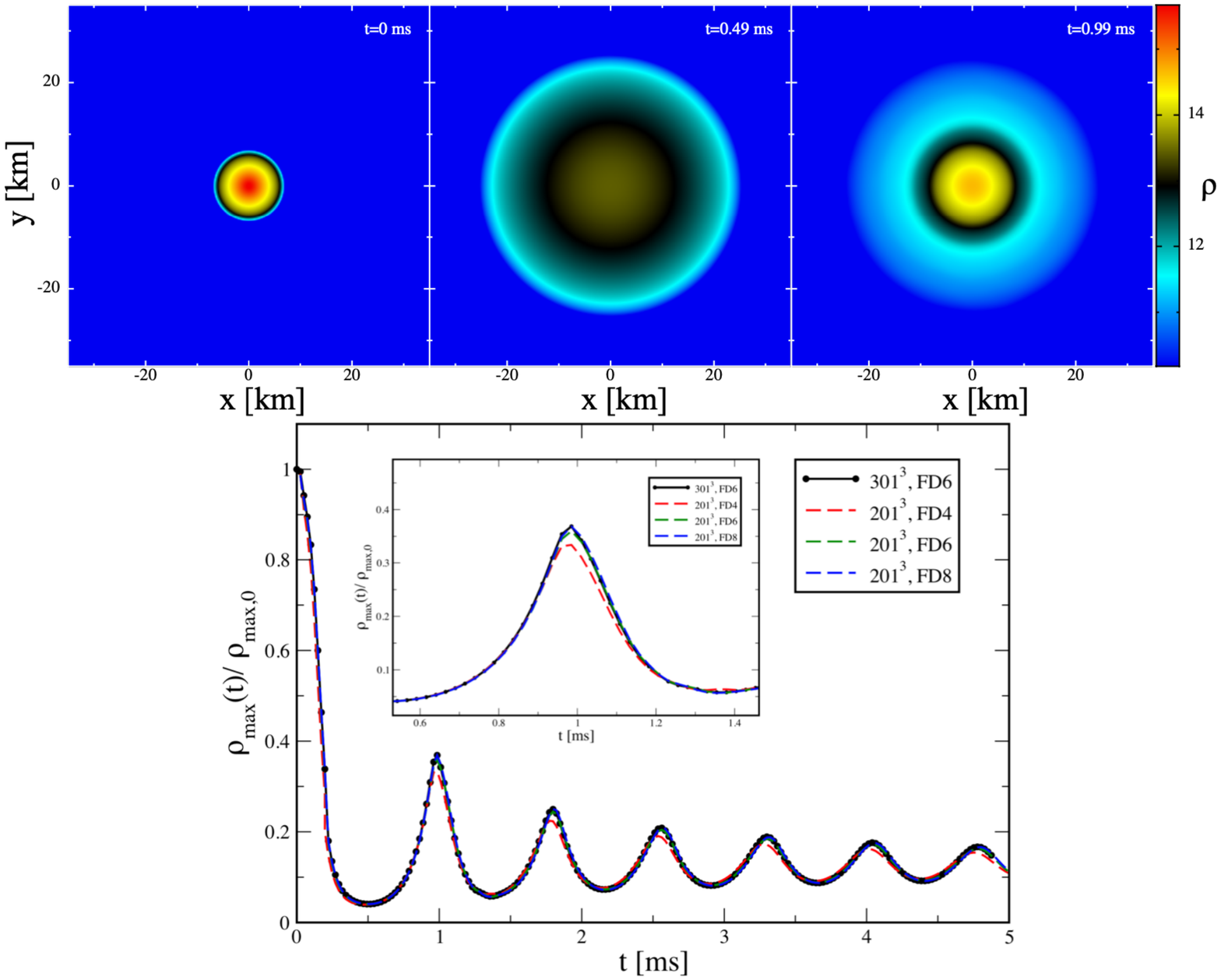}}
\vspace*{-0cm}
 \caption{Migration of a neutron star from the unstable to the stable branch. The star evolves, triggered by only truncation error,
 via violent oscillations ($v>0.5c$) towards the stable branch.The bottom panel shows the evolution of the central density for different
 spatial resolutions and Finite Differencing (FD) orders. Simulations performed with code \Sb.}
   \label{fig:migration}
\end{center}
\end{figure}

\begin{figure}
\begin{center}
\centerline{\includegraphics[width=\textwidth,angle=0]{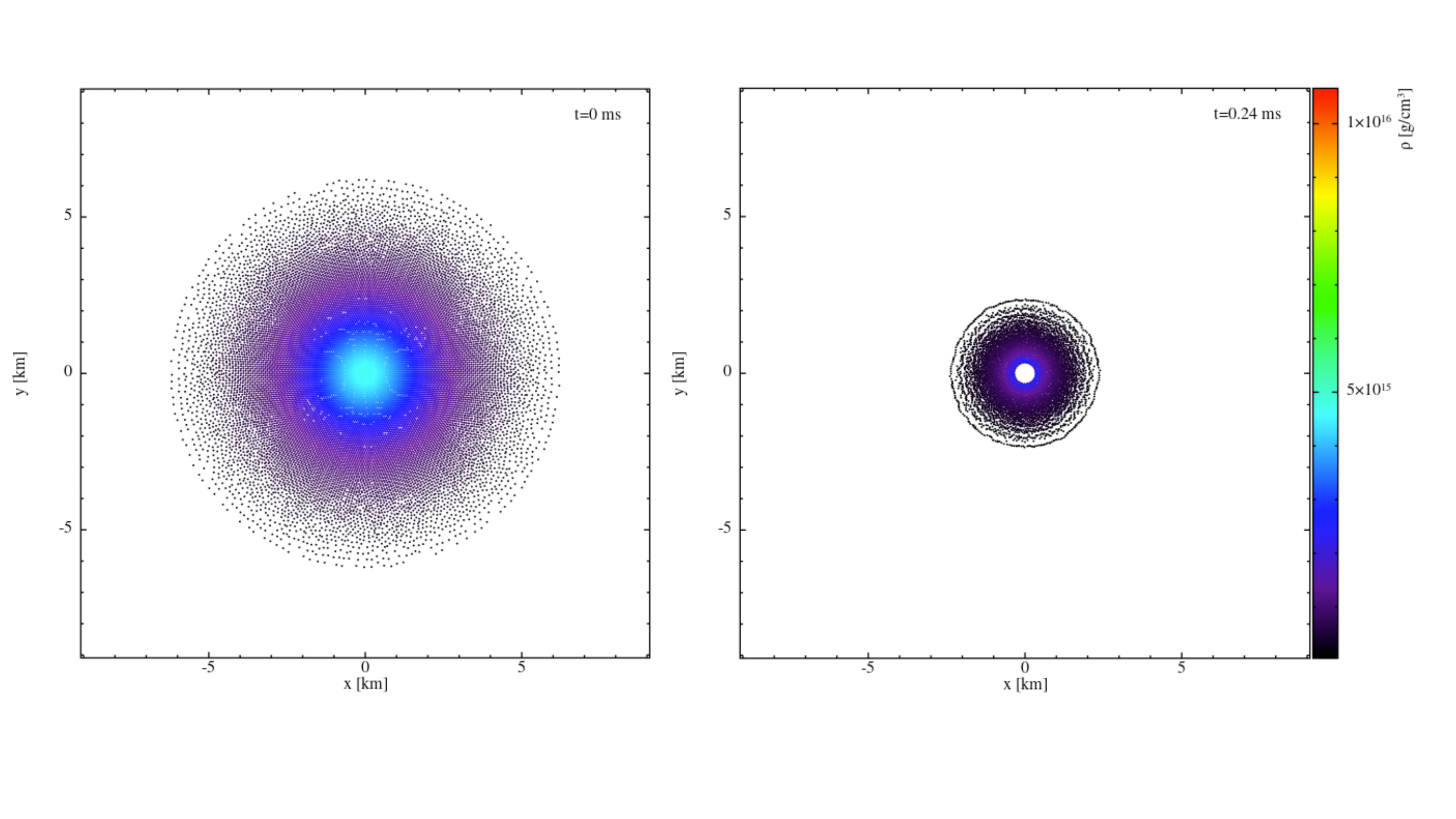}}
\vspace*{-1cm}
 \caption{Same star as in the previous figure, but this time a small inward velocity perturbation was applied. This is enough to trigger the collapse to a black hole. Shown are particle positions within $|z|< 0.7$ km. Simulations performed with code \Sb.}
   \label{fig:collapse}
\end{center}
\end{figure}

{\em Migration of an unstable neutron star to the stable branch}\\
A more complex test case involves an unstable initial configuration of a neutron star \cite[(Font et al. 2002, Cordero-Carrillon et al. 2009,
Bernuzzi \& Hilditch 2010)]{font02,cordero09,bernuzzi10}. According to Eulerian studies, the evolution depends delicately on the star's initial perturbation: if just evolved, the truncation error
alone drives the star to violent oscillations (with $v>0.5c$) and it finally settles on the stable branch. If, on the other hand, a small radial inward
velocity perturbation of only $\delta v_r= -0.005c \sin\left( \pi r /R \right)$ is applied, the star collapses and forms a black  hole. Can we confirm these results with \Sb?
Yes, we find again very close agreement with the Eulerian studies. Our results for the first case is shown in Fig.~\ref{fig:migration}. The upper panel
row shows different stages of the violent oscillation, the lower one shows the evolution of the central stellar density (for different numerical resolutions
and Finite Difference (FD) orders). When the small inward velocity perturbation is applied, the same star collapses to a black hole, see Fig.~\ref{fig:collapse}.

\section{Summary and conclusions}
\label{sec:summary}

In this paper we have described some of the recent developments related to SPH. Our focus
was on further improving SPH's accuracy without sacrificing its excellent conservation properties.
The new elements include high-order Wendland functions as SPH kernels, accurate gradients 
that require the inversion of a small matrix and new measures to steer dissipation in SPH. The first
of these measures is based on transferring Finite Volume techniques to SPH. More specifically, 
we perform slope-limited reconstructions between particle pairs and use these reconstructed values
in the artificial dissipation terms which massively reduces unnecessary dissipation even if the dissipation
parameters are kept at large, constant values. The results can be further improved by additionally 
making the dissipation parameters time dependent and steer them based on monitoring the exact 
conservation of entropy.\\
These new elements have been implemented into two codes that were developed from scratch: the Newtonian code \Ma (Rosswog 2020a)
and the fully general relativistic code \SB (Rosswog \& Diener 2021). Both codes have delivered 
results of very high accuracy and will be used in our future studies of astrophysical gas dynamics.

\newpage
\noindent{\bf Acknowledgements}\\
The author has been supported by the Swedish Research Council (VR) under grant number 2016\_03657,
by the Swedish National Space Board under grant number Dnr. 107/16, by the research environment grant 
"Gravitational Radiation and Electromagnetic Astrophysical Transients (GREAT)" funded by the 
Swedish Research council (VR) under Dnr 2016\_06012 and by the Knut and Alice Wallenberg Foundation 
under grant Dnr. KAW 2019.0112.  It is a great pleasure to acknowledge the very productive collaboration
with P. Diener, co-developer of \Sb.

\end{document}